\documentclass[preprint,aps,draft,pre,superscriptaddress,citeautoscript,floatfix,longbibliography]{revtex4-1}
\usepackage{graphics,graphicx,mathrsfs,color,amsmath,appendix,framed,tcolorbox,bm,enumerate}
\usepackage{empheq}

\newcommand{\ud}{\mathrm{d}}

\newcommand{\change}[1]{\textcolor{black}{#1}}

\begin{document}

\title{Capillary Imbibition in a Square Tube}
\author{Tian Yu}
\affiliation{Center of Soft Matter Physics and its Applications, Beihang University, Beijing 100191, China}
\affiliation{School of Physics and Nuclear Energy Engineering, Beihang University, Beijing 100191, China}
\author{Jiajia Zhou}
\email[]{jjzhou@buaa.edu.cn}
\affiliation{Center of Soft Matter Physics and its Applications, Beihang University, Beijing 100191, China}
\affiliation{School of Chemistry, Key Laboratory of Bio-Inspired Smart Interfacial Science and Technology of Ministry of Education, Beihang University, Beijing 100191, China}
\author{Masao Doi}
\email[]{masao.doi@buaa.edu.cn}
\affiliation{Center of Soft Matter Physics and its Applications, Beihang University, Beijing 100191, China}


\begin{abstract}
When a square tube is brought in contact with bulk liquid, the liquid wets the corners of the tube, and creates finger-like wetted region. 
The wetting of the liquid then takes place with the growth of two parts, the bulk part where the cross section is entirely filled with the liquid and the finger part where the cross section of the tube is partially filled. 
In the previous works, the growth of these two parts has been discussed separately. 
Here we conduct the analysis by explicitly accounting for the coupling of the two parts. 
We propose coupled equations for the liquid imbibition in both parts and show that 
(a) the length of each part, $h_0$ and  $h_1$, both increases in time $t$ following the Lucas-Washburn's law,  $h_0 \sim t^{1/2}$ and $h_1 \sim t^{1/2}$, 
but that (b) the coefficients are different from those obtained in the previous analysis which ignored the coupling. 
\end{abstract}

\maketitle


\section{Introduction}
The spontaneous filling of liquids into microchannels driven by capillary action is crucial for a broad range of applications, such as in microfluidic devices 
\cite{grunze1999driven, gau1999liquid, lai2010microchip}, 
lithography \cite{unger2000monolithic}, DNA manipulation \cite{burns1998integrated}, 
and liquid management in low-gravitation environments \cite{weislogel2003some}. 
A pioneering work on the dynamics of the imbibition was conducted by 
Lucas \cite{lucas1918time} and Washburn \cite{washburn1921dynamics} nearly 
a century ago.  
They considered the imbibition process in a circular tube 
which is brought in contact with bulk liquid. 
When the effect of gravity and inertia are ignored, they showed that the time 
dependence of the filling length, $h(t)$, is described by
\begin{equation}
\label{corrugate_1}
h(t) = k \, t^{1/2} = \sqrt {\frac{ \gamma a \cos\theta}{2\eta}} \, t^{1/2} \, ,
\end{equation}
where $a$ is the radius of the tube, $\theta$ is the equilibrium contact angle of the liquid to the tube surface, $\eta$ and $\gamma$  are the viscosity and surface tension of the liquid. 
The $t^{1/2}$ scaling of Lucas-Washburn has been confirmed experimentally in macroscopic systems as well as in nanoscale systems \cite{schebarchov2008dynamics, dimitrov2007capillary, YaoYang2017, YaoYang2018, YaoYang2018a}.

The Lucas-Washburn formula was obtained by considering the bulk
part of the liquid only, and ignoring the front part which involves
complex boundaries and complex flow fields. 
Such treatment is justified as long as the size of the front portion is small compared
with the bulk part.  
If the tube has a circular cross section, such condition is fulfilled since the liquid front takes a spherical shape and its length remains finite during the
imbibition process.  
On the other hand, if the tube has a triangular, or a square cross section (or 
in general polygonal shape), the effect of the front part cannot be neglected. 
In such a tube,  the liquid wets the corners, and forms ``fingers''. 
The liquid imbibition thus takes place with two parts, the bulk part and the finger part, both grow in time. 

Many studies have been performed for the capillary filling in non-circular tubes. 
Theoretical calculations have been done for the filling length $h(t)$  
in tubes having various cross sections,  such as triangular \cite{rye1996wetting, romero1996flow, mann1995flow}, rectangular \cite{ichikawa2004interface, yang2011dynamics}, and skewed U-shaped channels \cite{chen2009capillary}. 
Those calculations indicate that $h(t)$ obeys the Lucas-Washburn $t^{1/2}$ scaling. 
The front coefficient $k$ in Eq.~(\ref{corrugate_1}) varies depending on the geometry and the roughness of the channel walls \cite{ouali2013wetting}, and 
was shown consistently less than what Lucas-Washburn 
equation (\ref{corrugate_1}) predicts \cite{Chauvet2012}.
In all these calculations, the effect of the finger part has been ignored.

\change{The growth of the finger part has also been studied separately by a few groups.
Ransohoff and Radke \cite{ransohoff1988laminar} calculated the viscous resistance 
of the finger as a function of the surface shear viscosity and the contact angle. 
Dong and Chatzis \cite{dong1995imbibition} utilized this result to calculate the 
liquid imbibition in the corners of a square tube.
They also measured the advancement of the finger part by introducing several slugs of liquid into a square tube with both ends are sealed. 
In this case, the motion of the bulk can be ignored since the total amount of liquid is fixed. 
The experimental measurement were consistent with the theoretic predictions.  
Such studies have shown that the growth of the finger also obeys the Lucas-Washburn law (\ref{corrugate_1}).}

In this paper, we will consider the problem of liquid imbibition in a square tube, accounting for the coupling explicitly between the bulk part and the finger part. 
Similar system has been studied by Weislogel \cite{Weislogel2012} using the Laplacian scaling method \cite{Weislogel2008}. 
We shall show that both bulk part and the finger part grows in time obeying the Lucas-Washburn law, but differ in the numerical coefficients.
The imbibition is slowed down by ca 3\% due to the presence of the finger part.

\change{Gravity is another important factor in determining the imbibition dynamics. 
In the case of a horizontal tube when the imbibition direction is perpendicular to the direction of the gravity, the fingers may become unstable \cite{Manning2015, Rascon2016} when the tube size is larger than the capillary length. 
In the case of a vertical tube, the gravity influences the dynamics of the finger and bulk differently, because the mass involved in the bulk flow is much larger than that of finger flow. 
For the bulk part, the liquid in the tube eventually reached the Jurin's height and the imbibition stops. 
However, for the finger part, the tip of the finger is shown to move as $t^{1/3}$ \cite{Ponomarenko2011}, a different scaling to the classical Lucas-Washburn.
In this paper, we shall only consider a horizontal tube with a tube size smaller than the capillary length, therefore the effect of gravity can be neglected.}

\section{Model and Theory}

\begin{figure}[ht]
{\includegraphics[ scale=0.5,draft=false]{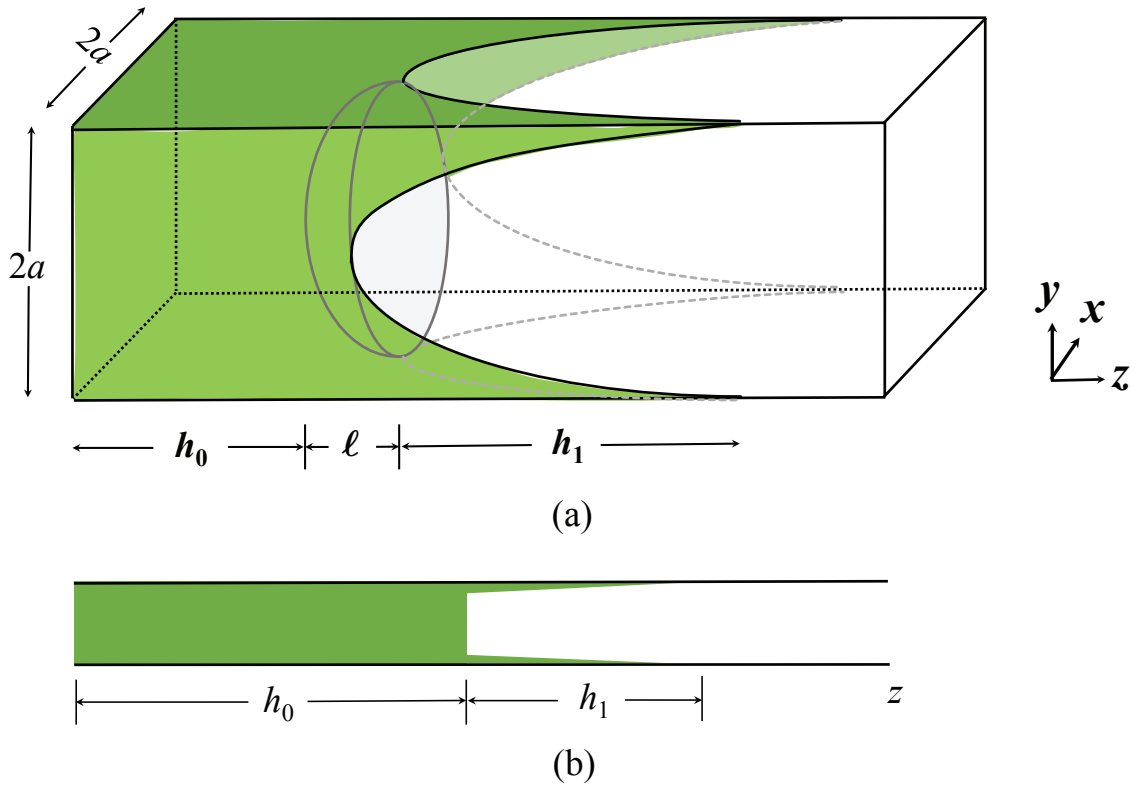}}
\caption{
Schematic picture of capillary imbibition of a liquid into a square tube. $(a)$ Perspective view -- The saturation $s(z)$ is equal to 1 for $0<z<h_0$, and decreases to $s^*$ in the transition region $h_0<z<h_0+\ell$. The present analysis for the finger part is valid for $z>h_0+\ell$. However, since $h_0$ and $h_1$ are much larger than $\ell$, we ignore the transition part and conduct the calculation assuming $\ell=0$. The saturation $s(z)$ decreases with $z$ for $z>h_0$ and goes to 0 at $z=h_0+h_1$, as illustrated in the side view $(b)$.  The side length of the square cross section is $2a$.} 
\label{fig:sketch}
\end{figure}

We consider a horizontal square tube with a side length $2a$, put in touch with an infinite reservoir of liquid of viscosity $\eta$ and surface tension $\gamma$. 
A schematic picture of the system is illustrated in Fig.~\ref{fig:sketch}(a).
Here we consider the simple case of fully wetted liquid, i.e., the contact angle $\theta$ is equal to zero.
A bulk liquid imbibes into the tube, accompanied by advancing fingers that wet the corners. 
The fingers can form when the liquid's contact angle is less than a critical contact angle $\theta_c = 45^{\circ}$ for a square tube \cite{concus1969behavior, langbein1990shape, higuera2008capillary}.
The imbibition length of liquid in the bulk is denoted by $h_0$ and the length of the meniscus front of the thin fingers is $h_1$.  
The liquid length of the transition region connecting the bulk and the thin fingers is $\ell$.
During the imbibition process, the length $\ell$ remains finite and becomes negligibly small at late times when both $h_0$ and $h_1$ become large. 
Therefore, we will consider a simplified picture shown in Fig.~\ref{fig:sketch}(b),  and neglect the transition region. 

We shall derive the time evolution equations for both the bulk and finger parts of the imbibing fluid using Onsager principle \cite{doi2013soft}.
The Onsager principle presents a general framework to derive the time evolution equation for non-equilibrium system, and the method has been successfully applied to various soft matter systems \cite{doi2015onsager, MengFanlong2016a, DiYana2016, XuXianmin2016, ManXingkun2016, ZhouJiajia2017, ManXingkun2017, DiYana2018}.
For the present problem, this principle amounts to the least energy dissipation principle in Stokesian hydrodynamics, which states that the dynamics of system can be directly determined by the minimum of the Rayleighian defined by
\begin{equation}
\label{corrugate_1_1}
\mathscr{R}=\dot{F}+\Phi\, ,     
\end{equation}
where $\dot{F}$ represents the time derivative of the free energy of  the system, and $\Phi$ represents the energy dissipation function, equal to half of the work done to the liquid per unit time.

\subsection{Free energy}

\begin{figure}[ht]
{\includegraphics[scale=0.5,draft=false]{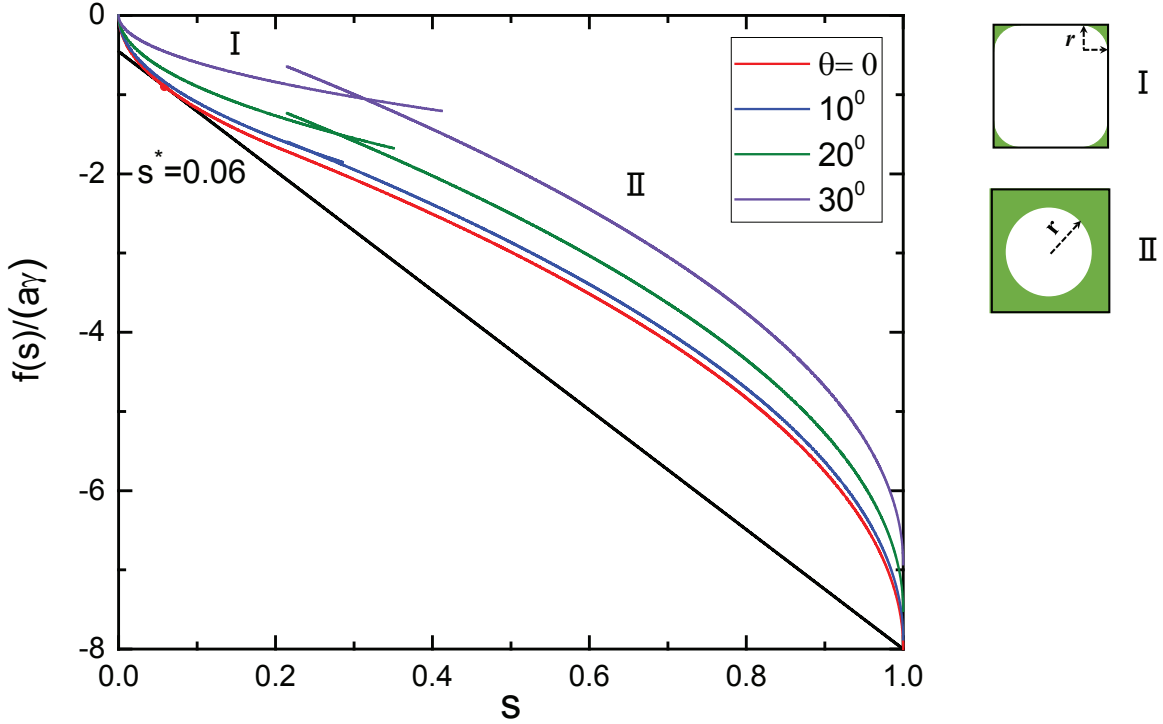}}
\caption{
\change{The free energy density is plotted as a function of the saturation $s$ for different contact angles $\theta$ by Eq.~(\ref{eq:fsO}) (case II, where there are no vapor-solid interfaces) and Eq.~(\ref{eq:fs1}) (case I, where vapor-solid interfaces exist and the liquid-vapor interfaces consist of four arcs near the corners).}    The black straight line goes through the two points with $s=s^*$ and $s=1$ of the curve of the free energy density for $\theta=0$. The slope of the straight line is equal to the derivative of the free energy density at the point of the equilibrium saturation $s^*$ through Eq.~(\ref{eq:fs1}).
}
\label{fig:freeE}
\end{figure}

 \change{In the plane perpendicular to the tube axis, the profile of the meniscus, i.e., the liquid-vapor interface, is a part of a circle (Fig. \ref{fig:freeE}).  
We define $r(z)$ as the radius of curvature of the liquid-vapor interface in the cross-section located at location $z$, } and saturation \change{$s(z)$} is the fraction of area occupied by the wetting liquid in a cross-section of the tube. 
The radius \change{$r(z)$ }can be written as a function of \change{$s(z)$}.
The free energy of the \change{whole} system can be written as 
\begin{equation}
  F = \int \ud z f(s(z)),
\end{equation}
where $f(s)$ is the local free energy density (free energy per unit length), which is a function of the local saturation $s(z)$. 
\change{We can distinguish two scenarios, characterized by whether the vapor-solid interface exists in the cross-section of the tube.}

\begin{itemize}
\item \change{When the inner surface of the tube is fully wetted by the liquid, there are no vapor-solid interface in the cross-section, and the liquid-vapor interface forms a closed circle.
This is illustrated in Fig. \ref{fig:freeE} case II.}
In this case, the saturation has to satisfy the condition 
\begin{equation}
  1 - \frac{\pi}{4} \le \change{s(z)} \le 1.
\end{equation}
The radius of the circle is given by
\begin{equation}
  \change{r(z)} = \sqrt{ \frac{4}{\pi} \big(1-\change{s(z)}\big) } a.
\end{equation}
The free energy density is given by 
\begin{equation}
  f(s) = 8 a (\gamma_{\rm LS} - \gamma_{\rm VS}) + 2\pi \change{r(z)} \gamma 
  = \left[ -8 \cos\theta + 4 \sqrt{\pi \big(1-\change{s(z)}\big)} \right] a \gamma .
  \label{eq:fsO}
\end{equation}         
where $\gamma_{\rm LS}$ and $\gamma_{\rm VS}$ are the interfacial tensions at the liquid-solid and vapor-solid interfaces, respectively. 
In the last line, we have used the Young's relation for the contact angle, $\gamma_{\rm LS} + \gamma \cos\theta = \gamma_{\rm VS}$. 

\item \change{When there are solid surfaces exposed to the air, as illustrated in Fig. \ref{fig:freeE} case I, the liquid-vapor interfaces consist of separated arcs in four corners. 
In this case, a critical saturation $s_c$ exists, above which the vapor-solid interfaces disappear. }
This critical saturation is given by
\begin{equation}
  s_c = \frac{C}{( \cos\theta - \sin\theta)^2}, \quad C = \cos^2 \theta - \sin\theta \cos\theta - (\frac{\pi}{4} - \theta). 
\end{equation}
The saturation has to satisfy the condition 
\begin{equation}
  0 \le \change{s(z)} \le s_c.
\end{equation}
The radius $r$ is given by 
\begin{equation}
  \change{r(z)} = \sqrt{ \frac{\change{s(z)}}{C} } a,
\end{equation}
and the free energy density is given by
\begin{eqnarray}
  f(s) &=& 8 \big(\change{r(z)} \cos\theta-\change{r(z)} \sin\theta \big)(\gamma_{\rm LS}-\gamma_{\rm VS}) + (2\pi-8\theta) \change{r(z)}\gamma 
 \nonumber \\
  &=& \big[-8 (\cos\theta-\sin\theta)\cos\theta + 2\pi-8\theta\big] a\gamma \sqrt{\frac{\change{s(z)}}{C}}. 
  \label{eq:fs1}
\end{eqnarray}

\end{itemize}

Figure \ref{fig:freeE} shows the free energy density as a function of the saturation $s$ for different contact angles $\theta$ [given by Eq. (\ref{eq:fsO}) and Eq. (\ref{eq:fs1})].
For contact angle $\theta > 0$, one can show that the critical saturation $s_c > 1 - \frac{\pi}{4}$, thus in the saturation range $1-\frac{\pi}{4} < \change{s(z)} < s_c$, both states are possible. 

For large saturation, the second derivative of the free energy density $\ud^2 f(s)/ \ud s^2$ is negative, indicating an unstable state. 
One of the possible stable state is $s=1$, the fully saturated state. 
The other possible state is given by drawing a straight line passing through $f(s=1)$ point, and the line is also tangential to the free energy curve at a small saturation $s=s^*$. 
This is shown as the black line for the case of $\theta=0$ in Fig.~\ref{fig:freeE}. 
For the perfectly wetting case ($\theta=0$), the saturation $s^* \simeq 0.06$ (See Appendix \ref{app:freeE} for details).

\subsection{Dissipation function}

We assume that liquid imbibes slowly in a horizontal capillary tube and ignore the effect of gravity and inertia.
The flow of liquid is almost one-dimensional, i.e., the $z$-component  of the flow velocity, $v_z$, is much larger than those in the other two directions. 
Thus the flow of liquid satisfies the Stokes equation
\begin{equation}
\label{corrugate_6}
 \eta \left( \frac{\partial^2 v_z}{\partial x^2}+\frac{\partial^2 v_z}{\partial y^2} \right) =\frac{\partial p}{\partial z}\,   ,       
\end{equation}
\change{where $\partial p/\partial z$ is the pressure gradient along the tube axis.}
The flow velocity $v_z$ can be expressed in  a dimensionless form as \cite{ransohoff1988laminar} 
\begin{equation}
\label{corrugate_7}
 \bar{u}=\frac{ \eta v_z}{\change{L}^2(-\frac{\partial p}{\partial z})}\,   ,      
\end{equation}
\change{where $L$ is a characteristic length scale. 
For the finger part, it is nature to use the radius of curvature $L=r(z)$, while for the bulk flow, the characteristic length is the tube size $L=a$.
In this way,  Eq.~(\ref{corrugate_6}) becomes a Poisson equation}
\begin{equation}
\label{corrugate_8}
\frac{\partial^2 \bar{u}}{\partial \bar{x}^2}+\frac{\partial^2 \bar{u}}{\partial \bar{y}^2} =-1\,   ,      
\end{equation}
with no-slip boundary conditions $\bar{u}=0$ at the liquid-solid interfaces and shear-free boundary condition $\bm{n}\cdot\ \nabla\bar{u}=0$ at the free surface, where $\bm{n}$ is the normal vector of the meniscus surface. 

The friction constant $\xi $ of the system is given by the Darcy's law
\begin{equation}
\label{corrugate_9}
\frac{\partial p}{\partial z}=-\xi Q\, ,     
\end{equation}
where $ Q=\int \mathrm{d}x\mathrm{d}y \, \change{v_z} $ is the volume flux of liquid threading a cross section per unit time. 
Combining Eq.~(\ref{corrugate_7}) and Eq.~(\ref{corrugate_9}), $\xi $  is written as
\begin{equation}
\label{corrugate_10}
\xi = \frac{\eta}{\change{L}^4 \int \mathrm{d}\bar{x}\mathrm{d}\bar{y}   \mathop{}\!\bar{u} }\, ,     
\end{equation}
\change{where the integral is obtained by solving the Poisson equation.
We performed numerical calculation using the finite element method in Matlab.}


For a perfectly wetting liquid in the bulk and the fingers, the friction constant $\xi$ can be calculated as a function of saturation $s$. 
For the bulk part $s=1$, the friction constant is
\begin{equation}
\label{corrugate_11}
     \xi (s=1)=\frac{\eta B_0}{a^4}, \quad  B_0 \simeq 1.7784.
\end{equation}
For the finger part
\begin{equation}
\label{corrugate_12}
     \xi (s<s^*)\change{=\frac{106.5\eta}{r(z)^4}}=\frac{\eta B_1}{a^4\change{s(z)}^2}, \quad B_1 \simeq 5.0 .
\end{equation}
\change{These values agree with the results of Refs. \cite{ouali2013wetting, ransohoff1988laminar}.}
The dissipation function is then given by
\label{corrugate_12_1}
\begin{equation}
  \Phi = \frac{1}{2} \int \ud z \, \xi(s) \, [Q(z)]^2. 
\end{equation}

\subsection{Capillary flow in the bulk}

We start with liquid flow in the bulk and ignore the effect of liquid imbibition in the fingers for the moment. 
Taking the case of perfectly wetting as an example, the time derivative of the free energy of the liquid in the bulk $\dot{F}$ is obtained through Eq.~(\ref{eq:fsO}) as
\begin{equation}
\label{corrugate_14}
\dot{F}=-8a\gamma\dot{h}_0\, ,     
\end{equation}
where $\dot{h}_0=\ud h_0 / \ud t$ denoting the advancing velocity of liquid in the bulk. The dissipation function $\Phi$ is given by 
\begin{equation}
\label{corrugate_15}
\Phi=\frac{1}{2}\xi Q^2 h_0\, ,     
\end{equation}
where the flux is
\begin{equation}
\label{corrugate_16}
Q=4a^2\dot{h}_0\, .    
\end{equation}
Substituting Eq.~(\ref{corrugate_16}) and  Eq.~(\ref{corrugate_11}) into Eq.~(\ref{corrugate_15}), it leads to
\begin{equation}
\label{corrugate_17}
\Phi=8B_0\eta h_0 \dot{h}_0^2 \, .    
\end{equation}
The evolution of $h_0(t)$ is determined by the minimum condition $\partial(\dot{F}+\Phi)/\partial \dot{h}_0=0$. 
This leads to
\begin{equation}
\label{corrugate_18}
h_0\dot{h}_0=\frac{a\gamma}{2B_0\eta} \, .    
\end{equation}
The above differential equation can be solved with the initial condition $h_0(t=0)=0$, and the result is 
\begin{equation}
\label{corrugate_19}
h_0=k_0\sqrt{\frac{a\gamma}{\eta}}t^{1/2} , \quad k_0 \simeq 0.75 \, .   
\end{equation}

\subsection{Capillary imbibition in the bulk and fingers}
We now consider the liquid flow and imbibition in the bulk and fingers simultaneously.
Through Eq. (\ref{eq:fsO}) and  Eq. (\ref{eq:fs1}) the free energy of liquid in the tube, including both the bulk and fingers, can be obtained as
\begin{equation}
\label{corrugate_20}
F=-8a\gamma h_0(t)-8 \alpha a\gamma\int_{h_0(t)}^{h_0(t)+h_1(t)}
\mathrm{d}z\sqrt{s(z,t)}\, ,
\end{equation}
where we have used a shorthand notation of $\alpha=\sqrt{1-\pi/4}$. 
The time derivative of the free energy $\dot{F}$  is written as
\begin{equation}
\label{corrugate_21}
\dot{F}=-8a\gamma  \dot{h}_0(t)+8\alpha a\gamma\sqrt{s^*} \dot{h}_0(t)-4\alpha a\gamma\int_{h_0(t)}^{h_0(t)+h_1(t)}\mathrm{d}z\bigg[\frac{\dot{s}(z,t)}{\sqrt{s(z,t)}}\bigg]\, . 
\end{equation}

For the liquid in the fingers, the volume conservation requires
\begin{equation}
\label{corrugate_22}
 \frac{\partial s}{\partial t}=-\frac{1}{4a^2}\frac{\partial Q_1}{\partial z}\,   ,       
\end{equation}
where $Q_1$ is the flux of liquid in fingers. 
Substituting Eq.~(\ref{corrugate_22}) into Eq.~(\ref{corrugate_21}), it leads to
\begin{equation}
\label{corrugate_23}
\dot{F}=-8a\gamma  \dot{h}_0(t)\Big[1-\alpha\sqrt{s^*}\Big]+\frac{\gamma\alpha }{2a}\int_{h_0(t)}^{h_0(t)+h_1(t)}\mathrm{d}z \bigg[\frac{\partial s}{\partial z}\bigg]s^{-\frac{3}{2}}Q_1- \frac {\alpha\gamma }{a}\frac {Q_1^* }{\sqrt{s^*}}\, ,
\end{equation}
where $Q_1^*$ is the flux at the entrance of the fingers ($z=h_0$), and the corresponding saturation $s^* \simeq 0.06$.

The conservation condition at the interface between the bulk part and the finger part, i.e., at $z=h_0(t)$, can be written as 
\begin{equation}
  \label{corrugate_25}
  Q_0 = Q_1^*+ 4a^2 (1-s^*)\dot h_0 \, .   
\end{equation}
The dissipation function $\Phi_0$ for liquid in the bulk is given by 
\begin{equation}
\label{corrugate_24}
\Phi_0=\frac{h_0}{2}\frac{B_0\eta}{a^4} Q_0^2 
= \frac{h_0}{2}\frac{B_0\eta}{a^4} \Big[Q_1^*+4a^2(1-s^*)\dot{h}_0\Big]^2 \, .    
\end{equation}

According to Eq.~(\ref{corrugate_12}) the dissipation function $\Phi_1$ for liquid in the fingers is written as
\begin{equation}
\label{corrugate_27}
\Phi_1=\frac{1}{2}\int_{h_0(t)}^{h_0(t)+h_1(t)}\mathrm{d}z \frac{\eta B_1}{a^4s^2(z,t)}Q_1^2(z,t)\, .    
\end{equation}
Through Eq.~(\ref{corrugate_24}) and  Eq.~(\ref{corrugate_27})  the total energy dissipation function $\Phi$ for liquid in the tube is given by
\begin{equation}
\label{corrugate_28}
\Phi=\frac{h_0}{2}\frac{B_0\eta}{a^4} \Big[Q_1^*+4a^2(1-s^*)\dot{h}_0\Big]^2 +\frac{1}{2}\int_{h_0(t)}^{h_0(t)+h_1(t)}\mathrm{d}z \frac{\eta B_1}{a^4s^2(z,t)}Q_1^2(z,t)\, .    
\end{equation}

The evolution equations for the imbibed liquid can be obtained by the variation $\delta (\dot{F}+\Phi)/ \delta \dot{h}_0=0$ and $\delta (\dot{F}+\Phi)/\delta Q_1=0$,

\begin{equation}
\label{corrugate_29}
\dot{h}_0=\frac{1}{4a^2(1-s^*)}\Big[\frac{2a^3\gamma \big(1-\alpha \sqrt{s^*} \big)}{B_0\eta(1-s^*)}\frac{1}{h_0}- Q_1^* \Big]\, ,
\end{equation}

\begin{equation}
\label{corrugate_30}
Q_1=-\frac{\alpha a^3\gamma }{2B_1\eta}s^{\frac{1}{2}}\frac{\partial s}{\partial z}\, .
\end{equation}
Substituting Eq.~(\ref{corrugate_22}) into Eq.~(\ref{corrugate_30}), it leads to the following time evolution equation of the fingers

\begin{equation}
\label{corrugate_31}
\frac{\partial s}{\partial t}=C_1\frac{\partial }{\partial z}\Big[s^{\frac{1}{2}}\frac{\partial s}{\partial z}\Big]\, ,
\end{equation}
where 
\begin{equation}
\label{corrugate_32}
C_1=\frac{\alpha a\gamma }{8B_1\eta}\, .
\end{equation}
Equation (\ref{corrugate_31}) has the same form with Dong's work \cite{dong1995imbibition}, which fixed the location of the entrance of the fingers and ignored the flow of liquid in the bulk. 

The evolution equation (\ref{corrugate_31}) for the fingers involves a moving boundary at $z=h_0(t)$, and it is difficult to solve numerically. 
We get around by the following variable transformation 
\begin{equation}
\label{corrugate_33}
z'=z-h_0,\quad  \tau=t\, .
\end{equation}
The evolution equation~(\ref{corrugate_31}) becomes
\begin{equation}
\label{corrugate_34}
\frac{\partial s}{\partial \tau}=C_1\frac{\partial }{\partial z'}\Big[s^{\frac{1}{2}}\frac{\partial s}{\partial z'}\Big]+\dot{h}_0 \frac{\partial s}{\partial z'}\, .
\end{equation}

Furthermore, we perform the  dimensionless transformation 
\begin{equation}
\label{corrugate_34_1}
\tilde{z'}=\frac{z'}{a},\quad  \tilde{h}_0=\frac{h_0}{a}\, ,
\end{equation}
\begin{equation}
\label{corrugate_34_2}
\tilde{\tau}=\frac{\tau}{a\eta/\gamma},\quad    \tilde{Q}_1=\frac{\eta}{\gamma}\frac{Q_1}{a^2}\, .
\end{equation}
The evolution equations (\ref{corrugate_29}), (\ref{corrugate_30}) and (\ref{corrugate_34}) take the following dimensionless forms
\begin{equation}
\label{corrugate_34_3}
\dot{\tilde{h}}_0=\frac{1}{4(1-s^*)}\Big[\frac{2\big(1-\alpha \sqrt{s^*} \big)}{B_0(1-s^*)}\frac{1}{\tilde{h}_0}- \tilde{Q}_1^* \Big]\, ,
\end{equation}
\begin{equation}
\label{corrugate_34_4}
\tilde{Q}_1=-\frac{\alpha}{2B_1}s^{\frac{1}{2}}\frac{\partial s}{\partial \tilde{z'}}\, .
\end{equation}
\begin{equation}
\label{corrugate_34_5}
\frac{\partial s}{\partial \tilde{\tau}}=\frac{\alpha}{8B_1}\frac{\partial }{\partial \tilde{z'}}\Big[s^{\frac{1}{2}}\frac{\partial s}{\partial \tilde{z'}}\Big]+\dot{\tilde{h}}_0 \frac{\partial s}{\partial \tilde{z'}}\, .
\end{equation}
The boundary conditions to Eq.~(\ref{corrugate_34_5}) are
\begin{equation}
\label{corrugate_35}
s|_{\tilde{z'}=0}=s^*=0.06,\qquad  s|_{\tilde{z'}\rightarrow \infty}=0  \, .
\end{equation}
Now the evolution of meniscus in the fingers  becomes a fixed boundaries problem and can be solved numerically using backward difference method. 

To summarize, the dynamics of the bulk $\tilde{h}_0(t)$ is given by an ordinary differential equation (\ref{corrugate_34_3}). 
It coupled to the finger flow through the flux $\tilde{Q}_1^*$ at the entrance of the finger. 
The dynamics of the finger part is given by a partial differential equation (\ref{corrugate_34_5}) with the boundary conditions (\ref{corrugate_35}). 
It cannot be solved without knowing the dynamics of the bulk part, because it contains the term $\dot{\tilde{h}}_0$. 
To obtain the full dynamics, one need to solve these two differential equations simultaneously, with suitable initial conditions.
\change{The initial conditions should be physically reasonable, for example, they have to be smooth and monotonically decreasing function from $s^*$ to 0. 
After a short period of time, different initial conditions all converge to the same profile, which leads to the same long-time dynamics.   
We elaborated the effect of initial conditions in Appendix \ref{app:ic}.}

\section{Results and discussion}

\begin{figure}[ht]
{\includegraphics[scale=0.5,draft=false]{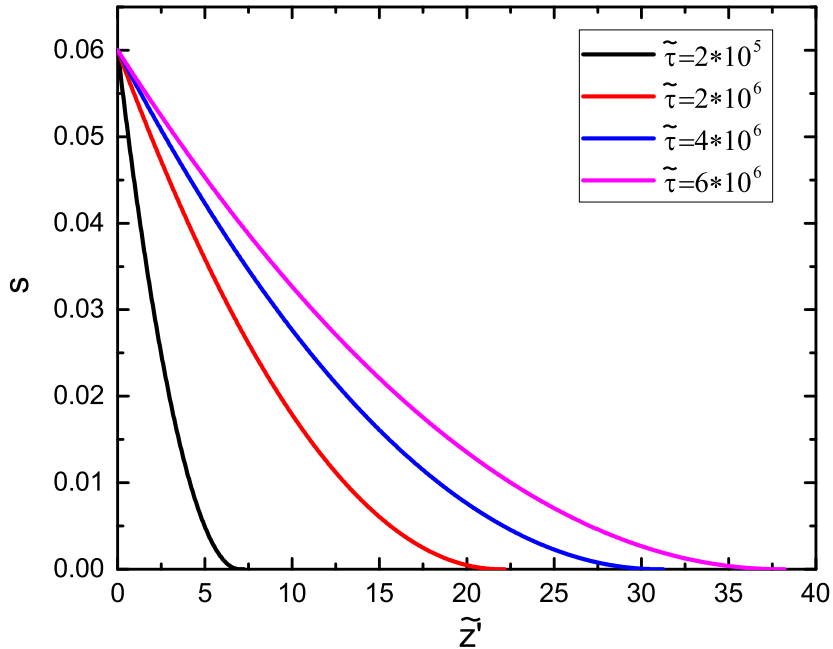}}
\caption{
The distributions of saturation $s$ with respect to  $\tilde{z'}$ for liquid in the fingers. Each curve corresponds to a different time $\tilde{\tau}$.}
\label{fig:finger}
\end{figure}

The ODE~(\ref{corrugate_34_3}) and PDE~(\ref{corrugate_34_5}) are a set of coupled equations describing the dynamic of liquid flow in the bulk and fingers. 
The corresponding numerical solution of saturation $s$ with respect to $\tilde{z'}$ for the fingers changing with time $\tilde{\tau}$ is shown in Fig.~\ref{fig:finger}. 
As time goes on, the saturation $s$ and the length of meniscus front both increase with $\tilde{\tau}$,  while the imbibition velocity of the meniscus front reduces qualitatively with $\tilde{\tau}$.

\begin{figure}[ht]
{\includegraphics[scale=0.5,draft=false]{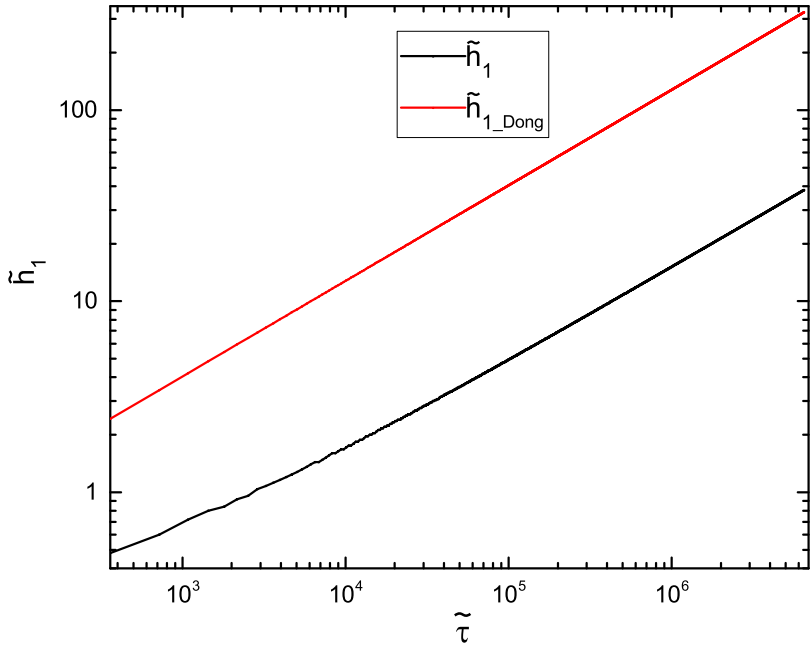}}
\caption{
The length of the fingers, $\tilde{h}_1$, is plotted against time, $\tilde{\tau}$. The black line is the numerical result based on Eq.~(\ref{corrugate_34_3}) and Eq.~(\ref{corrugate_34_5}), and the red line is the corresponding result from Ref.~\cite{dong1995imbibition}  where the movement of bulk flow is not considered.
}
\label{fig:h1}
\end{figure}

The length of finger, or the position of the tip of the finger with respect to the bulk fluid, $\tilde{h}_1$, is displayed in Fig.~\ref{fig:h1}.  
The black line is the numerical result of $\tilde{h}_1$ based on Eqs.~(\ref{corrugate_34_3}) and (\ref{corrugate_34_5}).
The solution can be nicely fitted with the power function
\begin{equation}
\label{corrugate_36}
\tilde{h}_1=k_1 \tilde{\tau}^{1/2} \, ,   
\end{equation}
with a spreading factor $k_1=0.015$. 
The red line in Fig.~\ref{fig:h1} is the analytic solution from Ref.~\cite{dong1995imbibition}, which considered only the evolution of liquid in the fingers while the bulk liquid does not move.
In this finger-only imbibition, the finger front follows Lucas-Washburn's $\tilde{\tau}^{1/2}$ scaling with a spreading factor $k_1=0.1278$, which is about ten times larger than our case. 
Through the comparison above, we know in practice the liquid flow in the bulk has a significant impact on meniscus evolution in the fingers and dramatically reduces the spreading factor of meniscus front of the fingers.


The evolution of liquid length in the bulk $\tilde{h}_0$ can be numerically calculated  based on Eq.~(\ref{corrugate_34_3}) and (\ref{corrugate_34_5}), taking into consideration of the coupling between the bulk and the fingers.
The solution can also well fitted by the power function
\begin{equation}
\label{corrugate_37}
\tilde{h}_0=k_0\tilde{\tau}^{1/2} \, ,   
\end{equation}
with a spreading factor $k_0=0.728$.  
Equation (\ref{corrugate_37}) indicates the time evolution of liquid in the bulk also follows the Lucas-Washburn's law with a reduced spreading factor compared with that in Eq.~(\ref{corrugate_19}), $k_0=0.75$, which ignores the effect of the fingers. 
Hence, for liquid flow and imbibition in the horizontal square tube, the coupling between the bulk and the fingers decreases the spreading factor of each other, which is intuitively embodied in the second terms on the right-hand of Eq.~(\ref{corrugate_34_3}) and Eq.~(\ref{corrugate_34_5}), respectively. 
A similar result has been presented in a vertical square tube by Bico and Qu{\'e}r{\'e} \cite{bico2002rise}, who found the measured equilibrium rise height values of liquid in the bulk are around 6\% smaller than theoretical predictions and attributed this reduction to the wetting fingers by Princen Model \cite{princen1969capillary_1, princen1969capillary_2}. 

Since both the bulk and the fingers follow Lucas-Washburn's $t^{1/2}$ scaling law, the ratio between the finger length and the bulk length remains constant
\begin{equation}
  \frac{\tilde{h}_1}{\tilde{h}_0} = \frac{k_1}{k_2} \simeq 0.0206.
\end{equation}
This ratio is smaller than the value 0.02959 predicted by the Laplacian scaling method \cite{Weislogel2012}.  



\section{Conclusions} 
In this paper, we have studied the capillary imbibition and flow of a liquid along the corners of a horizontal square tube. 
The spontaneous filling is composed of two parts: the liquid in the bulk and the leading meniscus of the fingers. 
We first presented the dynamics of liquid in the bulk ignoring the effect of the fingers, which is the standard Lucas-Washburn result.
Then we proposed a set of coupled differential equations to describe the evolution of liquid in the bulk and in the fingers.
We solved the equations numerically. 
The main results of our study are
\begin{enumerate}
\item[(1)] The time evolution of both the bulk ($\tilde{h}_0$) and the fingers ($\tilde{h}_1$) follow the Lucas-Washburn's classical $t^{1/2}$ scaling law. 

\item[(2)] The spreading factors $k_0=0.728$ and $k_1=0.015$ of the bulk and fingers based on coupling effect are lower than that those predicted by models considered only one of them in isolation, especially for $k_1$, the value of which will be reduced to an order of magnitude compared with Ref.~\cite{dong1995imbibition}.
\end{enumerate}

Our numerical results indicate that Lucas-Washburn's $t^{1/2}$ scaling is robust to predict the evolution of liquid in a square tube in the viscous regime, and the coupling effect plays an important role in determining the spreading factors.

\begin{acknowledgments}
This work was supported by the National Natural Science Foundation of China (NSFC) through the Grant Nos. 21504004 and 21774004. 
M.D. acknowledges the financial support of the Chinese Central Government in the Thousand Talents Program. 
\end{acknowledgments}

\appendix

\section{Calculation of $s^*$}
\label{app:freeE}

The free energy density at fully saturation is given by Eq.~(\ref{eq:fsO})
\begin{equation}
  f(1)=-8 \cos\theta a \gamma .
\end{equation}

The free energy density at small saturation is given by Eq.~(\ref{eq:fs1})
\begin{eqnarray}
  f(s) &=& g(\theta) \sqrt{s} a \gamma, \\
  g(\theta) &=& \frac{ - 8 (\cos\theta - \sin\theta) \cos\theta + 2\pi - 8\theta }
  { [ \cos^2 \theta - \sin\theta \cos\theta - (\pi/4-\theta) ]^{1/2}} .
\end{eqnarray}

The equilibrium between the fully saturation case and partial saturation case is given by the condition
\begin{equation}
  \frac{ \ud f(s)}{\ud s} \Big|_{s=s^*} = \frac{ f(1)-f(s^*)}{1-s^*}.
\end{equation}
This corresponds to a straight line passing through $s=s^*$ and $s=1$ points, while also tangential to the free energy curve at $s=s^*$ (see Fig.~\ref{fig:freeE}). 

For the fully wetting case $\theta=0$, this condition becomes
\begin{equation}
  \frac{ - 8 + 8 \alpha \sqrt{s^*} }{1-s^*} = - 4 \alpha \frac{1}{\sqrt{s^*}},
\end{equation}
where $\alpha=\sqrt{1-\pi/4}$. 
Solving for $s^*$ and one gets
\begin{equation}
  s^* = \left[ \frac{1-\sqrt{1-\alpha^2}}{\alpha} \right]^2 \simeq 0.0603.
\end{equation}

\section{Effect of initial conditions }
\label{app:ic}

\begin{figure}[ht]
(a) {\includegraphics[scale=0.5,draft=false]{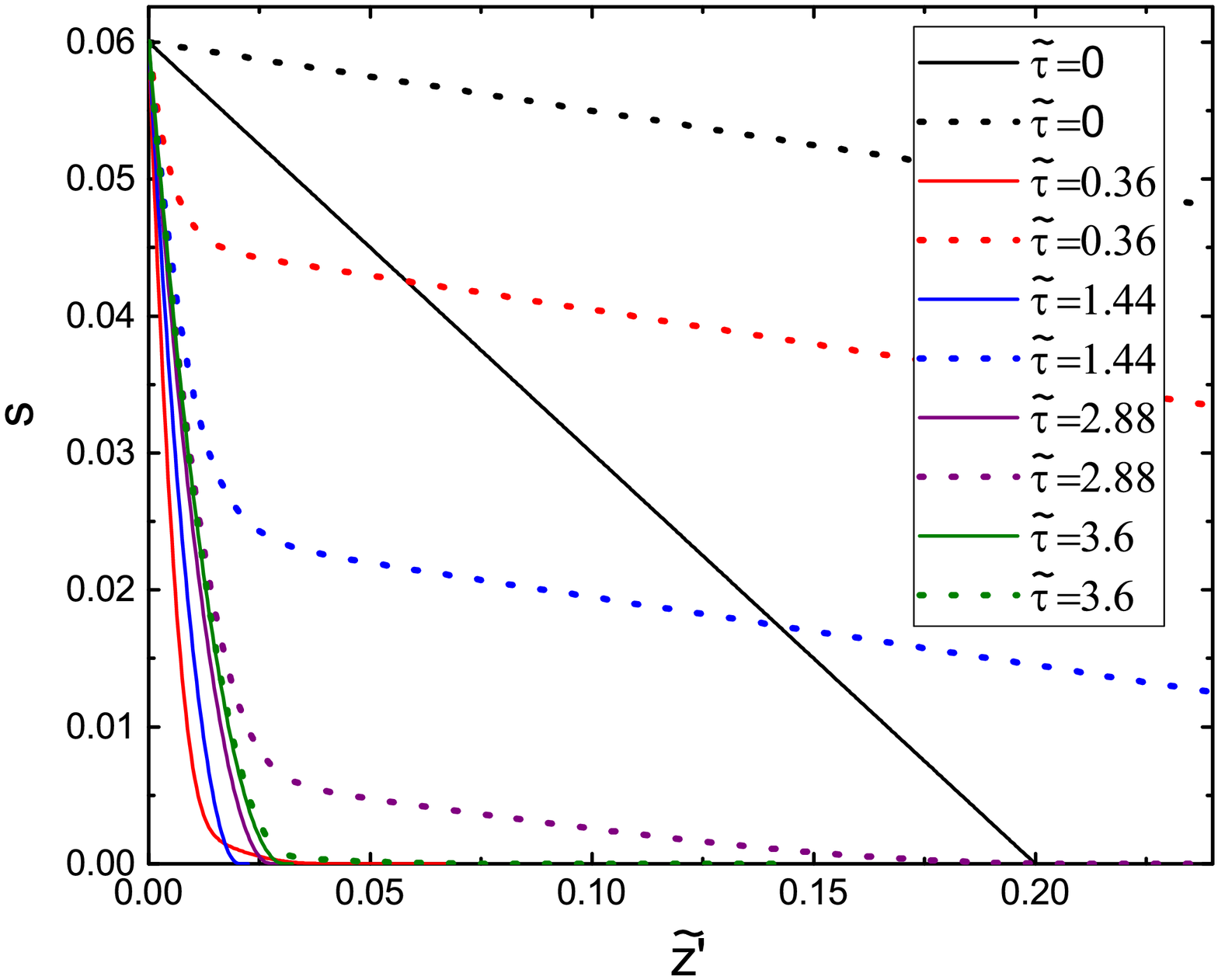}} \\
(b) {\includegraphics[scale=0.5,draft=false]{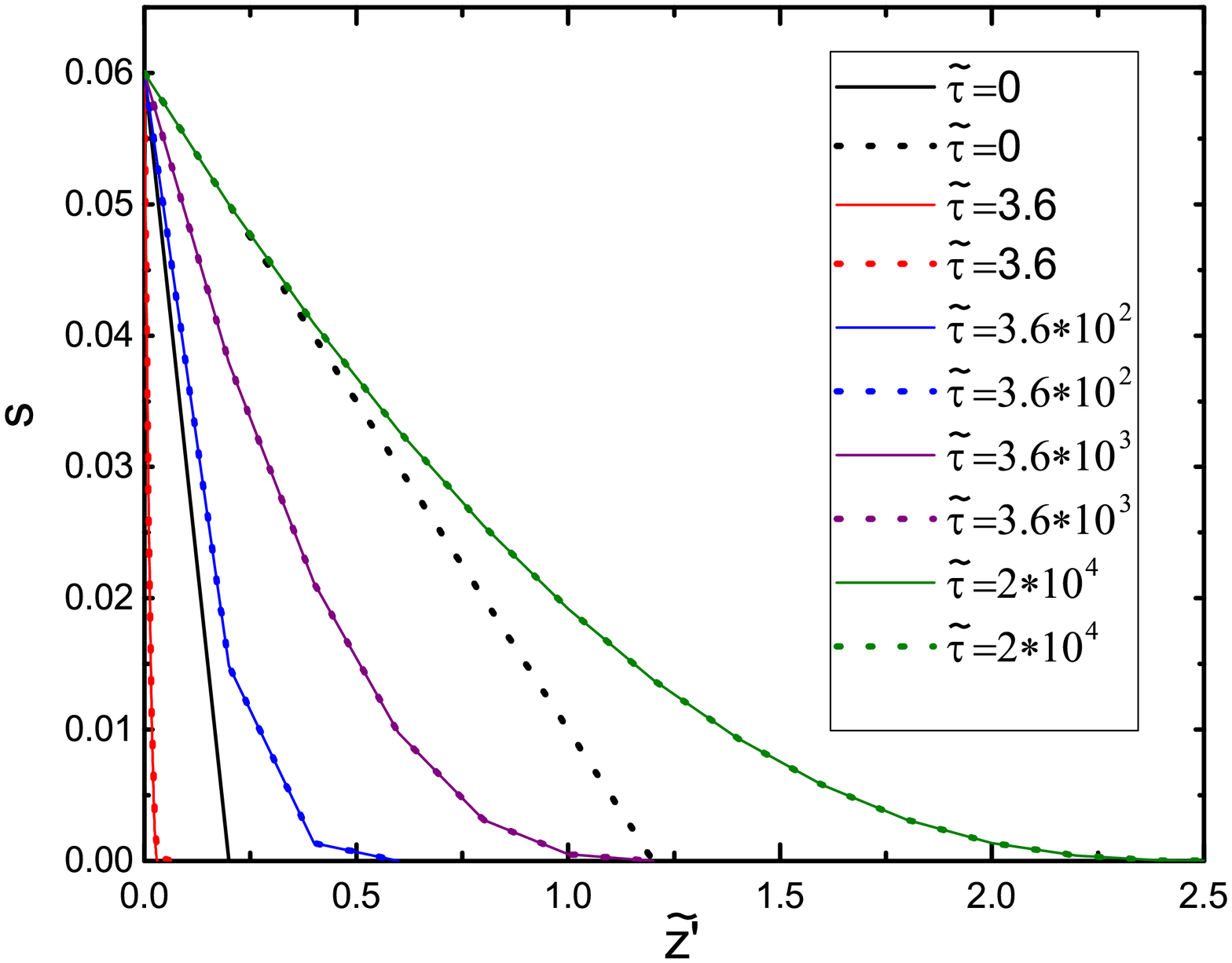}}
\caption{
The evolution of the saturation $s$ of the fingers part for different initial conditions. The solid lines are the evolution results based on the initial condition (I) $s=-(0.06/0.2) \tilde{z'}+0.06$, and  the dot lines are the evolution results based on (II) $s=-(0.06/1.2) \tilde{z'}+0.06$. Both of the calculations are based the same  parameter values: $\tilde{h}_{0}(\tilde{\tau}=0)=0.2$. \change{Short-time dynamics is shown in (a) and long-time dynamics is shown in (b).}
}
\label{fig:ic}
\end{figure}

Here we analyze the effect of initial conditions on the numerical results of Eq.~(\ref{corrugate_34_3}) and Eq.~(\ref{corrugate_34_5}).
\change{The discrete steps for the time and the position are $\Delta \tilde{\tau}=3.6 \times 10^{-6}$, $\Delta \tilde{z'}=4\times 10^{-4}$.} 
For the bulk part, the initial length of the bulk is $\tilde{h}_{0}(\tilde{\tau}=0)=0.2$. 
For the fingers part, we give two different initial profiles of the saturation $s$ for comparison, and both of them are linear forms: 
\begin{itemize}
\item[(I)] $s=-(0.06/0.2) \tilde{z'}+0.06$. 
The corresponding initial values for $\tilde{h}_1$ and $\tilde{Q}_1^*$ are $\tilde{h}_{1}(\tilde{\tau}=0)=0.2$, $\tilde{Q}_{1}^*(\tilde{\tau}=0)=0.0034$.
\item[(II)] $s=-(0.06/1.2) \tilde{z'}+0.06$.
The corresponding initial values for $\tilde{h}_1$ and $\tilde{Q}_1^*$ are $\tilde{h}_{1}(\tilde{\tau}=0)=1.2$, $\tilde{Q}_{1}^*(\tilde{\tau}=0)=0.00057$. 
\end{itemize}

\change{The evolution of the saturation $s$ for the fingers part are plotted in Fig.~\ref{fig:ic} for (a) short-time and (b) long-time. 
One can see that after a relatively short time $\tilde{\tau}=3.6$, the profiles from two different initial conditions converge to almost the same profile. 
After $\tilde{\tau}=3.6$, the two profiles are indistinguishable and exhibit the same long-time dynamics.
Therefore, the initial conditions only affect the dynamics in a very short time scale, and have no influence on the long-time dynamics of the meniscus.}

\bibliography{yu}

\begin{thebibliography}{43}%
\makeatletter
\providecommand \@ifxundefined [1]{%
 \@ifx{#1\undefined}
}%
\providecommand \@ifnum [1]{%
 \ifnum #1\expandafter \@firstoftwo
 \else \expandafter \@secondoftwo
 \fi
}%
\providecommand \@ifx [1]{%
 \ifx #1\expandafter \@firstoftwo
 \else \expandafter \@secondoftwo
 \fi
}%
\providecommand \natexlab [1]{#1}%
\providecommand \enquote  [1]{``#1''}%
\providecommand \bibnamefont  [1]{#1}%
\providecommand \bibfnamefont [1]{#1}%
\providecommand \citenamefont [1]{#1}%
\providecommand \href@noop [0]{\@secondoftwo}%
\providecommand \href [0]{\begingroup \@sanitize@url \@href}%
\providecommand \@href[1]{\@@startlink{#1}\@@href}%
\providecommand \@@href[1]{\endgroup#1\@@endlink}%
\providecommand \@sanitize@url [0]{\catcode `\\12\catcode `\$12\catcode
  `\&12\catcode `\#12\catcode `\^12\catcode `\_12\catcode `\%12\relax}%
\providecommand \@@startlink[1]{}%
\providecommand \@@endlink[0]{}%
\providecommand \url  [0]{\begingroup\@sanitize@url \@url }%
\providecommand \@url [1]{\endgroup\@href {#1}{\urlprefix }}%
\providecommand \urlprefix  [0]{URL }%
\providecommand \Eprint [0]{\href }%
\providecommand \doibase [0]{http://dx.doi.org/}%
\providecommand \selectlanguage [0]{\@gobble}%
\providecommand \bibinfo  [0]{\@secondoftwo}%
\providecommand \bibfield  [0]{\@secondoftwo}%
\providecommand \translation [1]{[#1]}%
\providecommand \BibitemOpen [0]{}%
\providecommand \bibitemStop [0]{}%
\providecommand \bibitemNoStop [0]{.\EOS\space}%
\providecommand \EOS [0]{\spacefactor3000\relax}%
\providecommand \BibitemShut  [1]{\csname bibitem#1\endcsname}%
\let\auto@bib@innerbib\@empty
\bibitem [{\citenamefont {Grunze}(1999)}]{grunze1999driven}%
  \BibitemOpen
  \bibfield  {author} {\bibinfo {author} {\bibfnamefont {Michael}\ \bibnamefont
  {Grunze}},\ }\bibfield  {title} {\enquote {\bibinfo {title} {Driven
  liquids},}\ }\href@noop {} {\bibfield  {journal} {\bibinfo  {journal}
  {Science}\ }\textbf {\bibinfo {volume} {283}},\ \bibinfo {pages} {41--42}
  (\bibinfo {year} {1999})}\BibitemShut {NoStop}%
\bibitem [{\citenamefont {Gau}\ \emph {et~al.}(1999)\citenamefont {Gau},
  \citenamefont {Herminghaus}, \citenamefont {Lenz},\ and\ \citenamefont
  {Lipowsky}}]{gau1999liquid}%
  \BibitemOpen
  \bibfield  {author} {\bibinfo {author} {\bibfnamefont {Hartmut}\ \bibnamefont
  {Gau}}, \bibinfo {author} {\bibfnamefont {Stephan}\ \bibnamefont
  {Herminghaus}}, \bibinfo {author} {\bibfnamefont {Peter}\ \bibnamefont
  {Lenz}}, \ and\ \bibinfo {author} {\bibfnamefont {Reinhard}\ \bibnamefont
  {Lipowsky}},\ }\bibfield  {title} {\enquote {\bibinfo {title} {Liquid
  morphologies on structured surfaces: from microchannels to microchips},}\
  }\href@noop {} {\bibfield  {journal} {\bibinfo  {journal} {Science}\ }\textbf
  {\bibinfo {volume} {283}},\ \bibinfo {pages} {46--49} (\bibinfo {year}
  {1999})}\BibitemShut {NoStop}%
\bibitem [{\citenamefont {Lai}\ \emph {et~al.}(2010)\citenamefont {Lai},
  \citenamefont {Yang},\ and\ \citenamefont {Shieh}}]{lai2010microchip}%
  \BibitemOpen
  \bibfield  {author} {\bibinfo {author} {\bibfnamefont {Yu-Hsuan}\
  \bibnamefont {Lai}}, \bibinfo {author} {\bibfnamefont {Jing-Tang}\
  \bibnamefont {Yang}}, \ and\ \bibinfo {author} {\bibfnamefont {Dar-Bin}\
  \bibnamefont {Shieh}},\ }\bibfield  {title} {\enquote {\bibinfo {title} {A
  microchip fabricated with a vapor-diffusion self-assembled-monolayer method
  to transport droplets across superhydrophobic to hydrophilic surfaces},}\
  }\href@noop {} {\bibfield  {journal} {\bibinfo  {journal} {Lab on a Chip}\
  }\textbf {\bibinfo {volume} {10}},\ \bibinfo {pages} {499--504} (\bibinfo
  {year} {2010})}\BibitemShut {NoStop}%
\bibitem [{\citenamefont {Unger}\ \emph {et~al.}(2000)\citenamefont {Unger},
  \citenamefont {Chou}, \citenamefont {Thorsen}, \citenamefont {Scherer},\ and\
  \citenamefont {Quake}}]{unger2000monolithic}%
  \BibitemOpen
  \bibfield  {author} {\bibinfo {author} {\bibfnamefont {Marc~A}\ \bibnamefont
  {Unger}}, \bibinfo {author} {\bibfnamefont {Hou-Pu}\ \bibnamefont {Chou}},
  \bibinfo {author} {\bibfnamefont {Todd}\ \bibnamefont {Thorsen}}, \bibinfo
  {author} {\bibfnamefont {Axel}\ \bibnamefont {Scherer}}, \ and\ \bibinfo
  {author} {\bibfnamefont {Stephen~R}\ \bibnamefont {Quake}},\ }\bibfield
  {title} {\enquote {\bibinfo {title} {Monolithic microfabricated valves and
  pumps by multilayer soft lithography},}\ }\href@noop {} {\bibfield  {journal}
  {\bibinfo  {journal} {Science}\ }\textbf {\bibinfo {volume} {288}},\ \bibinfo
  {pages} {113--116} (\bibinfo {year} {2000})}\BibitemShut {NoStop}%
\bibitem [{\citenamefont {Burns}\ \emph {et~al.}(1998)\citenamefont {Burns},
  \citenamefont {Johnson}, \citenamefont {Brahmasandra}, \citenamefont
  {Handique}, \citenamefont {Webster}, \citenamefont {Krishnan}, \citenamefont
  {Sammarco}, \citenamefont {Man}, \citenamefont {Jones}, \citenamefont
  {Heldsinger} \emph {et~al.}}]{burns1998integrated}%
  \BibitemOpen
  \bibfield  {author} {\bibinfo {author} {\bibfnamefont {Mark~A}\ \bibnamefont
  {Burns}}, \bibinfo {author} {\bibfnamefont {Brian~N}\ \bibnamefont
  {Johnson}}, \bibinfo {author} {\bibfnamefont {Sundaresh~N}\ \bibnamefont
  {Brahmasandra}}, \bibinfo {author} {\bibfnamefont {Kalyan}\ \bibnamefont
  {Handique}}, \bibinfo {author} {\bibfnamefont {James~R}\ \bibnamefont
  {Webster}}, \bibinfo {author} {\bibfnamefont {Madhavi}\ \bibnamefont
  {Krishnan}}, \bibinfo {author} {\bibfnamefont {Timothy~S}\ \bibnamefont
  {Sammarco}}, \bibinfo {author} {\bibfnamefont {Piu~M}\ \bibnamefont {Man}},
  \bibinfo {author} {\bibfnamefont {Darren}\ \bibnamefont {Jones}}, \bibinfo
  {author} {\bibfnamefont {Dylan}\ \bibnamefont {Heldsinger}},  \emph
  {et~al.},\ }\bibfield  {title} {\enquote {\bibinfo {title} {An integrated
  nanoliter dna analysis device},}\ }\href@noop {} {\bibfield  {journal}
  {\bibinfo  {journal} {Science}\ }\textbf {\bibinfo {volume} {282}},\ \bibinfo
  {pages} {484--487} (\bibinfo {year} {1998})}\BibitemShut {NoStop}%
\bibitem [{\citenamefont {Weislogel}(2003)}]{weislogel2003some}%
  \BibitemOpen
  \bibfield  {author} {\bibinfo {author} {\bibfnamefont {M.~M.}\ \bibnamefont
  {Weislogel}},\ }\bibfield  {title} {\enquote {\bibinfo {title} {Some
  analytical tools for fluids management in space: Isothermal capillary flows
  along interior corners},}\ }\href@noop {} {\bibfield  {journal} {\bibinfo
  {journal} {Advances in Space Research}\ }\textbf {\bibinfo {volume} {32}},\
  \bibinfo {pages} {163--170} (\bibinfo {year} {2003})}\BibitemShut {NoStop}%
\bibitem [{\citenamefont {Lucas}(1918)}]{lucas1918time}%
  \BibitemOpen
  \bibfield  {author} {\bibinfo {author} {\bibfnamefont {R.}~\bibnamefont
  {Lucas}},\ }\bibfield  {title} {\enquote {\bibinfo {title} {The time law of
  the capillary rise of liquids},}\ }\href@noop {} {\bibfield  {journal}
  {\bibinfo  {journal} {Kolloid-Zeitschrift}\ }\textbf {\bibinfo {volume}
  {23}},\ \bibinfo {pages} {15--22} (\bibinfo {year} {1918})}\BibitemShut
  {NoStop}%
\bibitem [{\citenamefont {Washburn}(1921)}]{washburn1921dynamics}%
  \BibitemOpen
  \bibfield  {author} {\bibinfo {author} {\bibfnamefont {E.~W.}\ \bibnamefont
  {Washburn}},\ }\bibfield  {title} {\enquote {\bibinfo {title} {The dynamics
  of capillary flow},}\ }\href@noop {} {\bibfield  {journal} {\bibinfo
  {journal} {Phys. Rev.}\ }\textbf {\bibinfo {volume} {17}},\ \bibinfo {pages}
  {273} (\bibinfo {year} {1921})}\BibitemShut {NoStop}%
\bibitem [{\citenamefont {Schebarchov}\ and\ \citenamefont
  {Hendy}(2008)}]{schebarchov2008dynamics}%
  \BibitemOpen
  \bibfield  {author} {\bibinfo {author} {\bibfnamefont {D.}~\bibnamefont
  {Schebarchov}}\ and\ \bibinfo {author} {\bibfnamefont {S.~C.}\ \bibnamefont
  {Hendy}},\ }\bibfield  {title} {\enquote {\bibinfo {title} {Dynamics of
  capillary absorption of droplets by carbon nanotubes},}\ }\href@noop {}
  {\bibfield  {journal} {\bibinfo  {journal} {Phys. Rev. E}\ }\textbf {\bibinfo
  {volume} {78}},\ \bibinfo {pages} {046309} (\bibinfo {year}
  {2008})}\BibitemShut {NoStop}%
\bibitem [{\citenamefont {Dimitrov}\ \emph {et~al.}(2007)\citenamefont
  {Dimitrov}, \citenamefont {Milchev},\ and\ \citenamefont
  {Binder}}]{dimitrov2007capillary}%
  \BibitemOpen
  \bibfield  {author} {\bibinfo {author} {\bibfnamefont {D.}~\bibnamefont
  {Dimitrov}}, \bibinfo {author} {\bibfnamefont {A.}~\bibnamefont {Milchev}}, \
  and\ \bibinfo {author} {\bibfnamefont {K.}~\bibnamefont {Binder}},\
  }\bibfield  {title} {\enquote {\bibinfo {title} {Capillary rise in nanopores:
  molecular dynamics evidence for the lucas-washburn equation},}\ }\href@noop
  {} {\bibfield  {journal} {\bibinfo  {journal} {Phys. Rev. Lett.}\ }\textbf
  {\bibinfo {volume} {99}},\ \bibinfo {pages} {054501} (\bibinfo {year}
  {2007})}\BibitemShut {NoStop}%
\bibitem [{\citenamefont {Yao}\ \emph {et~al.}(2017)\citenamefont {Yao},
  \citenamefont {Alexandris}, \citenamefont {Henrich}, \citenamefont
  {Auernhammer}, \citenamefont {Steinhart}, \citenamefont {Butt},\ and\
  \citenamefont {Floudas}}]{YaoYang2017}%
  \BibitemOpen
  \bibfield  {author} {\bibinfo {author} {\bibfnamefont {Yang}\ \bibnamefont
  {Yao}}, \bibinfo {author} {\bibfnamefont {Stelios}\ \bibnamefont
  {Alexandris}}, \bibinfo {author} {\bibfnamefont {Franziska}\ \bibnamefont
  {Henrich}}, \bibinfo {author} {\bibfnamefont {G\"{u}nter}\ \bibnamefont
  {Auernhammer}}, \bibinfo {author} {\bibfnamefont {Martin}\ \bibnamefont
  {Steinhart}}, \bibinfo {author} {\bibfnamefont {Hans-J\"{u}rgen}\
  \bibnamefont {Butt}}, \ and\ \bibinfo {author} {\bibfnamefont {George}\
  \bibnamefont {Floudas}},\ }\bibfield  {title} {\enquote {\bibinfo {title}
  {Complex dynamics of capillary imbibition of poly(ethylene oxide) melts in
  nanoporous alumina},}\ }\href {\doibase 10.1063/1.4978298} {\bibfield
  {journal} {\bibinfo  {journal} {J. Chem. Phys.}\ }\textbf {\bibinfo {volume}
  {146}},\ \bibinfo {pages} {203320} (\bibinfo {year} {2017})}\BibitemShut
  {NoStop}%
\bibitem [{\citenamefont {Yao}\ \emph {et~al.}(2018{\natexlab{a}})\citenamefont
  {Yao}, \citenamefont {Butt}, \citenamefont {Zhou}, \citenamefont {Doi},\ and\
  \citenamefont {Floudas}}]{YaoYang2018}%
  \BibitemOpen
  \bibfield  {author} {\bibinfo {author} {\bibfnamefont {Yang}\ \bibnamefont
  {Yao}}, \bibinfo {author} {\bibfnamefont {Hans-J{\"u}rgen}\ \bibnamefont
  {Butt}}, \bibinfo {author} {\bibfnamefont {Jiajia}\ \bibnamefont {Zhou}},
  \bibinfo {author} {\bibfnamefont {Masao}\ \bibnamefont {Doi}}, \ and\
  \bibinfo {author} {\bibfnamefont {George}\ \bibnamefont {Floudas}},\
  }\bibfield  {title} {\enquote {\bibinfo {title} {Capillary imibibition of
  polymer mixtures in nanopores},}\ }\href {\doibase
  10.1021/acs.macromol.7b02724} {\bibfield  {journal} {\bibinfo  {journal}
  {Macromolecules}\ }\textbf {\bibinfo {volume} {51}},\ \bibinfo {pages}
  {3059--3065} (\bibinfo {year} {2018}{\natexlab{a}})}\BibitemShut {NoStop}%
\bibitem [{\citenamefont {Yao}\ \emph {et~al.}(2018{\natexlab{b}})\citenamefont
  {Yao}, \citenamefont {Butt}, \citenamefont {Floudas}, \citenamefont {Zhou},\
  and\ \citenamefont {Doi}}]{YaoYang2018a}%
  \BibitemOpen
  \bibfield  {author} {\bibinfo {author} {\bibfnamefont {Yang}\ \bibnamefont
  {Yao}}, \bibinfo {author} {\bibfnamefont {Hans-J{\"u}rgen}\ \bibnamefont
  {Butt}}, \bibinfo {author} {\bibfnamefont {George}\ \bibnamefont {Floudas}},
  \bibinfo {author} {\bibfnamefont {Jiajia}\ \bibnamefont {Zhou}}, \ and\
  \bibinfo {author} {\bibfnamefont {Masao}\ \bibnamefont {Doi}},\ }\bibfield
  {title} {\enquote {\bibinfo {title} {Theory on capillary filling of polymer
  metls in nanpores},}\ }\href {\doibase 10.1002/marc.201800087} {\bibfield
  {journal} {\bibinfo  {journal} {Macromol. Rapid Commun.}\ } (\bibinfo {year}
  {2018}{\natexlab{b}}),\ 10.1002/marc.201800087}\BibitemShut {NoStop}%
\bibitem [{\citenamefont {Rye}\ \emph {et~al.}(1996)\citenamefont {Rye},
  \citenamefont {Yost},\ and\ \citenamefont {Mann}}]{rye1996wetting}%
  \BibitemOpen
  \bibfield  {author} {\bibinfo {author} {\bibfnamefont {R.~R.}\ \bibnamefont
  {Rye}}, \bibinfo {author} {\bibfnamefont {F.~G.}\ \bibnamefont {Yost}}, \
  and\ \bibinfo {author} {\bibfnamefont {J.~A.}\ \bibnamefont {Mann}},\
  }\bibfield  {title} {\enquote {\bibinfo {title} {Wetting kinetics in surface
  capillary grooves},}\ }\href@noop {} {\bibfield  {journal} {\bibinfo
  {journal} {Langmuir}\ }\textbf {\bibinfo {volume} {12}},\ \bibinfo {pages}
  {4625--4627} (\bibinfo {year} {1996})}\BibitemShut {NoStop}%
\bibitem [{\citenamefont {Romero}\ and\ \citenamefont
  {Yost}(1996)}]{romero1996flow}%
  \BibitemOpen
  \bibfield  {author} {\bibinfo {author} {\bibfnamefont {L.~A.}\ \bibnamefont
  {Romero}}\ and\ \bibinfo {author} {\bibfnamefont {F.~G.}\ \bibnamefont
  {Yost}},\ }\bibfield  {title} {\enquote {\bibinfo {title} {Flow in an open
  channel capillary},}\ }\href@noop {} {\bibfield  {journal} {\bibinfo
  {journal} {J. Fluid Mech.}\ }\textbf {\bibinfo {volume} {322}},\ \bibinfo
  {pages} {109--129} (\bibinfo {year} {1996})}\BibitemShut {NoStop}%
\bibitem [{\citenamefont {Mann~Jr}\ \emph {et~al.}(1995)\citenamefont
  {Mann~Jr}, \citenamefont {Romero}, \citenamefont {Rye},\ and\ \citenamefont
  {Yost}}]{mann1995flow}%
  \BibitemOpen
  \bibfield  {author} {\bibinfo {author} {\bibfnamefont {J.~A.}\ \bibnamefont
  {Mann~Jr}}, \bibinfo {author} {\bibfnamefont {L.}~\bibnamefont {Romero}},
  \bibinfo {author} {\bibfnamefont {R.~R.}\ \bibnamefont {Rye}}, \ and\
  \bibinfo {author} {\bibfnamefont {F.~G.}\ \bibnamefont {Yost}},\ }\bibfield
  {title} {\enquote {\bibinfo {title} {Flow of simple liquids down narrow {ssV}
  grooves},}\ }\href@noop {} {\bibfield  {journal} {\bibinfo  {journal} {Phys.
  Rev. E}\ }\textbf {\bibinfo {volume} {52}},\ \bibinfo {pages} {3967}
  (\bibinfo {year} {1995})}\BibitemShut {NoStop}%
\bibitem [{\citenamefont {Ichikawa}\ \emph {et~al.}(2004)\citenamefont
  {Ichikawa}, \citenamefont {Hosokawa},\ and\ \citenamefont
  {Maeda}}]{ichikawa2004interface}%
  \BibitemOpen
  \bibfield  {author} {\bibinfo {author} {\bibfnamefont {Naoki}\ \bibnamefont
  {Ichikawa}}, \bibinfo {author} {\bibfnamefont {Kazuo}\ \bibnamefont
  {Hosokawa}}, \ and\ \bibinfo {author} {\bibfnamefont {Ryutaro}\ \bibnamefont
  {Maeda}},\ }\bibfield  {title} {\enquote {\bibinfo {title} {Interface motion
  of capillary-driven flow in rectangular microchannel},}\ }\href@noop {}
  {\bibfield  {journal} {\bibinfo  {journal} {J. Colloid Interface Sci.}\
  }\textbf {\bibinfo {volume} {280}},\ \bibinfo {pages} {155--164} (\bibinfo
  {year} {2004})}\BibitemShut {NoStop}%
\bibitem [{\citenamefont {Yang}\ \emph {et~al.}(2011)\citenamefont {Yang},
  \citenamefont {Krasowska}, \citenamefont {Priest}, \citenamefont {Popescu},\
  and\ \citenamefont {Ralston}}]{yang2011dynamics}%
  \BibitemOpen
  \bibfield  {author} {\bibinfo {author} {\bibfnamefont {Die}\ \bibnamefont
  {Yang}}, \bibinfo {author} {\bibfnamefont {Marta}\ \bibnamefont {Krasowska}},
  \bibinfo {author} {\bibfnamefont {Craig}\ \bibnamefont {Priest}}, \bibinfo
  {author} {\bibfnamefont {Mihail~N}\ \bibnamefont {Popescu}}, \ and\ \bibinfo
  {author} {\bibfnamefont {John}\ \bibnamefont {Ralston}},\ }\bibfield  {title}
  {\enquote {\bibinfo {title} {Dynamics of capillary-driven flow in open
  microchannels},}\ }\href@noop {} {\bibfield  {journal} {\bibinfo  {journal}
  {J. Phys. Chem. C}\ }\textbf {\bibinfo {volume} {115}},\ \bibinfo {pages}
  {18761--18769} (\bibinfo {year} {2011})}\BibitemShut {NoStop}%
\bibitem [{\citenamefont {Chen}\ \emph {et~al.}(2009)\citenamefont {Chen},
  \citenamefont {Melvin}, \citenamefont {Rodriguez}, \citenamefont {Bell},\
  and\ \citenamefont {Weislogel}}]{chen2009capillary}%
  \BibitemOpen
  \bibfield  {author} {\bibinfo {author} {\bibfnamefont {Yongkang}\
  \bibnamefont {Chen}}, \bibinfo {author} {\bibfnamefont {Lawrence~S}\
  \bibnamefont {Melvin}}, \bibinfo {author} {\bibfnamefont {Santiago}\
  \bibnamefont {Rodriguez}}, \bibinfo {author} {\bibfnamefont {Donald}\
  \bibnamefont {Bell}}, \ and\ \bibinfo {author} {\bibfnamefont {Mark~M}\
  \bibnamefont {Weislogel}},\ }\bibfield  {title} {\enquote {\bibinfo {title}
  {Capillary driven flow in micro scale surface structures},}\ }\href@noop {}
  {\bibfield  {journal} {\bibinfo  {journal} {Microelectronic Engineering}\
  }\textbf {\bibinfo {volume} {86}},\ \bibinfo {pages} {1317--1320} (\bibinfo
  {year} {2009})}\BibitemShut {NoStop}%
\bibitem [{\citenamefont {Ouali}\ \emph {et~al.}(2013)\citenamefont {Ouali},
  \citenamefont {McHale}, \citenamefont {Javed}, \citenamefont {Trabi},
  \citenamefont {Shirtcliffe},\ and\ \citenamefont
  {Newton}}]{ouali2013wetting}%
  \BibitemOpen
  \bibfield  {author} {\bibinfo {author} {\bibfnamefont {F.~Fouzia}\
  \bibnamefont {Ouali}}, \bibinfo {author} {\bibfnamefont {Glen}\ \bibnamefont
  {McHale}}, \bibinfo {author} {\bibfnamefont {Haadi}\ \bibnamefont {Javed}},
  \bibinfo {author} {\bibfnamefont {Christophe}\ \bibnamefont {Trabi}},
  \bibinfo {author} {\bibfnamefont {Neil~J.}\ \bibnamefont {Shirtcliffe}}, \
  and\ \bibinfo {author} {\bibfnamefont {Michael~I.}\ \bibnamefont {Newton}},\
  }\bibfield  {title} {\enquote {\bibinfo {title} {Wetting considerations in
  capillary rise and imbibition in closed square tubes and open rectangular
  cross-section channels},}\ }\href {\doibase 10.1007/s10404-013-1145-5}
  {\bibfield  {journal} {\bibinfo  {journal} {Microfluid. Nanofluid.}\ }\textbf
  {\bibinfo {volume} {15}},\ \bibinfo {pages} {309--326} (\bibinfo {year}
  {2013})}\BibitemShut {NoStop}%
\bibitem [{\citenamefont {Chauvet}\ \emph {et~al.}(2012)\citenamefont
  {Chauvet}, \citenamefont {Geoffroy}, \citenamefont {Hamoumi}, \citenamefont
  {Prat},\ and\ \citenamefont {Joseph}}]{Chauvet2012}%
  \BibitemOpen
  \bibfield  {author} {\bibinfo {author} {\bibfnamefont {Fabien}\ \bibnamefont
  {Chauvet}}, \bibinfo {author} {\bibfnamefont {Sandrine}\ \bibnamefont
  {Geoffroy}}, \bibinfo {author} {\bibfnamefont {Abdelkrim}\ \bibnamefont
  {Hamoumi}}, \bibinfo {author} {\bibfnamefont {Marc}\ \bibnamefont {Prat}}, \
  and\ \bibinfo {author} {\bibfnamefont {Pierre}\ \bibnamefont {Joseph}},\
  }\bibfield  {title} {\enquote {\bibinfo {title} {Roles of gas in capillary
  filling of nanoslits},}\ }\href {\doibase 10.1039/c2sm25982f} {\bibfield
  {journal} {\bibinfo  {journal} {Soft Matter}\ }\textbf {\bibinfo {volume}
  {8}},\ \bibinfo {pages} {10738} (\bibinfo {year} {2012})}\BibitemShut
  {NoStop}%
\bibitem [{\citenamefont {Ransohoff}\ and\ \citenamefont
  {Radke}(1988)}]{ransohoff1988laminar}%
  \BibitemOpen
  \bibfield  {author} {\bibinfo {author} {\bibfnamefont {T.~C.}\ \bibnamefont
  {Ransohoff}}\ and\ \bibinfo {author} {\bibfnamefont {C.~J.}\ \bibnamefont
  {Radke}},\ }\bibfield  {title} {\enquote {\bibinfo {title} {Laminar flow of a
  wetting liquid along the corners of a predominantly gas-occupied noncircular
  pore},}\ }\href@noop {} {\bibfield  {journal} {\bibinfo  {journal} {J.
  Colloid Interface Sci.}\ }\textbf {\bibinfo {volume} {121}},\ \bibinfo
  {pages} {392--401} (\bibinfo {year} {1988})}\BibitemShut {NoStop}%
\bibitem [{\citenamefont {Dong}\ and\ \citenamefont
  {Chatzis}(1995)}]{dong1995imbibition}%
  \BibitemOpen
  \bibfield  {author} {\bibinfo {author} {\bibfnamefont {M.}~\bibnamefont
  {Dong}}\ and\ \bibinfo {author} {\bibfnamefont {I.}~\bibnamefont {Chatzis}},\
  }\bibfield  {title} {\enquote {\bibinfo {title} {The imbibition and flow of a
  wetting liquid along the corners of a square capillary tube},}\ }\href@noop
  {} {\bibfield  {journal} {\bibinfo  {journal} {J. Colloid Interface Sci.}\
  }\textbf {\bibinfo {volume} {172}},\ \bibinfo {pages} {278--288} (\bibinfo
  {year} {1995})}\BibitemShut {NoStop}%
\bibitem [{\citenamefont {Weislogel}(2012)}]{Weislogel2012}%
  \BibitemOpen
  \bibfield  {author} {\bibinfo {author} {\bibfnamefont {M.~M.}\ \bibnamefont
  {Weislogel}},\ }\bibfield  {title} {\enquote {\bibinfo {title} {Compound
  capillary rise},}\ }\href@noop {} {\bibfield  {journal} {\bibinfo  {journal}
  {J. Fluid Mech.}\ }\textbf {\bibinfo {volume} {709}},\ \bibinfo {pages}
  {622--647} (\bibinfo {year} {2012})}\BibitemShut {NoStop}%
\bibitem [{\citenamefont {Weislogel}\ \emph {et~al.}(2008)\citenamefont
  {Weislogel}, \citenamefont {Chen},\ and\ \citenamefont
  {Bolleddula}}]{Weislogel2008}%
  \BibitemOpen
  \bibfield  {author} {\bibinfo {author} {\bibfnamefont {M.~M.}\ \bibnamefont
  {Weislogel}}, \bibinfo {author} {\bibfnamefont {Y.}~\bibnamefont {Chen}}, \
  and\ \bibinfo {author} {\bibfnamefont {D.}~\bibnamefont {Bolleddula}},\
  }\bibfield  {title} {\enquote {\bibinfo {title} {A better
  nondimensionalization scheme for slender laminar flows: The laplacian
  operator scaling method},}\ }\href {\doibase 10.1063/1.2973900} {\bibfield
  {journal} {\bibinfo  {journal} {Phys. Fluids}\ }\textbf {\bibinfo {volume}
  {20}},\ \bibinfo {pages} {093602} (\bibinfo {year} {2008})}\BibitemShut
  {NoStop}%
\bibitem [{\citenamefont {Manning}\ and\ \citenamefont
  {Collicott}(2015)}]{Manning2015}%
  \BibitemOpen
  \bibfield  {author} {\bibinfo {author} {\bibfnamefont {Robert~E.}\
  \bibnamefont {Manning}}\ and\ \bibinfo {author} {\bibfnamefont {Steven~H.}\
  \bibnamefont {Collicott}},\ }\bibfield  {title} {\enquote {\bibinfo {title}
  {Existence of static capillary plugs in horizontal rectangular cylinders},}\
  }\href {\doibase 10.1007/s10404-015-1632-y} {\bibfield  {journal} {\bibinfo
  {journal} {Microfluid. Nanofluid.}\ }\textbf {\bibinfo {volume} {19}},\
  \bibinfo {pages} {1159--1168} (\bibinfo {year} {2015})}\BibitemShut {NoStop}%
\bibitem [{\citenamefont {Rasc{\'{o}}n}\ \emph {et~al.}(2016)\citenamefont
  {Rasc{\'{o}}n}, \citenamefont {Parry},\ and\ \citenamefont
  {Aarts}}]{Rascon2016}%
  \BibitemOpen
  \bibfield  {author} {\bibinfo {author} {\bibfnamefont {Carlos}\ \bibnamefont
  {Rasc{\'{o}}n}}, \bibinfo {author} {\bibfnamefont {Andrew~O.}\ \bibnamefont
  {Parry}}, \ and\ \bibinfo {author} {\bibfnamefont {Dirk G. A.~L.}\
  \bibnamefont {Aarts}},\ }\bibfield  {title} {\enquote {\bibinfo {title}
  {Geometry-induced capillary emptying},}\ }\href {\doibase
  10.1073/pnas.1606217113} {\bibfield  {journal} {\bibinfo  {journal} {{PNAS}}\
  }\textbf {\bibinfo {volume} {113}},\ \bibinfo {pages} {12633--12636}
  (\bibinfo {year} {2016})}\BibitemShut {NoStop}%
\bibitem [{\citenamefont {Ponomarenko}\ \emph {et~al.}(2011)\citenamefont
  {Ponomarenko}, \citenamefont {Qu{\'e}r{\'e}},\ and\ \citenamefont
  {Clanet}}]{Ponomarenko2011}%
  \BibitemOpen
  \bibfield  {author} {\bibinfo {author} {\bibfnamefont {Alexandre}\
  \bibnamefont {Ponomarenko}}, \bibinfo {author} {\bibfnamefont {David}\
  \bibnamefont {Qu{\'e}r{\'e}}}, \ and\ \bibinfo {author} {\bibfnamefont
  {Christophe}\ \bibnamefont {Clanet}},\ }\bibfield  {title} {\enquote
  {\bibinfo {title} {A universal law for capillary rise in corners},}\
  }\href@noop {} {\bibfield  {journal} {\bibinfo  {journal} {J. Fluid Mech.}\
  }\textbf {\bibinfo {volume} {666}},\ \bibinfo {pages} {146--154} (\bibinfo
  {year} {2011})}\BibitemShut {NoStop}%
\bibitem [{\citenamefont {Concus}\ and\ \citenamefont
  {Finn}(1969)}]{concus1969behavior}%
  \BibitemOpen
  \bibfield  {author} {\bibinfo {author} {\bibfnamefont {Paul}\ \bibnamefont
  {Concus}}\ and\ \bibinfo {author} {\bibfnamefont {Robert}\ \bibnamefont
  {Finn}},\ }\bibfield  {title} {\enquote {\bibinfo {title} {On the behavior of
  a capillary surface in a wedge},}\ }\href@noop {} {\bibfield  {journal}
  {\bibinfo  {journal} {{PNAS}}\ }\textbf {\bibinfo {volume} {63}},\ \bibinfo
  {pages} {292} (\bibinfo {year} {1969})}\BibitemShut {NoStop}%
\bibitem [{\citenamefont {Langbein}(1990)}]{langbein1990shape}%
  \BibitemOpen
  \bibfield  {author} {\bibinfo {author} {\bibfnamefont {Dieter}\ \bibnamefont
  {Langbein}},\ }\bibfield  {title} {\enquote {\bibinfo {title} {The shape and
  stability of liquid menisci at solid edges},}\ }\href@noop {} {\bibfield
  {journal} {\bibinfo  {journal} {J. Fluid Mech.}\ }\textbf {\bibinfo {volume}
  {213}},\ \bibinfo {pages} {251--265} (\bibinfo {year} {1990})}\BibitemShut
  {NoStop}%
\bibitem [{\citenamefont {Higuera}\ \emph {et~al.}(2008)\citenamefont
  {Higuera}, \citenamefont {Medina},\ and\ \citenamefont
  {Linan}}]{higuera2008capillary}%
  \BibitemOpen
  \bibfield  {author} {\bibinfo {author} {\bibfnamefont {F.~J.}\ \bibnamefont
  {Higuera}}, \bibinfo {author} {\bibfnamefont {A.}~\bibnamefont {Medina}}, \
  and\ \bibinfo {author} {\bibfnamefont {A.}~\bibnamefont {Linan}},\ }\bibfield
   {title} {\enquote {\bibinfo {title} {Capillary rise of a liquid between two
  vertical plates making a small angle},}\ }\href@noop {} {\bibfield  {journal}
  {\bibinfo  {journal} {Phys. Fluids}\ }\textbf {\bibinfo {volume} {20}},\
  \bibinfo {pages} {102102} (\bibinfo {year} {2008})}\BibitemShut {NoStop}%
\bibitem [{\citenamefont {Doi}(2013)}]{doi2013soft}%
  \BibitemOpen
  \bibfield  {author} {\bibinfo {author} {\bibfnamefont {Masao}\ \bibnamefont
  {Doi}},\ }\href@noop {} {\emph {\bibinfo {title} {Soft Matter Physics}}}\
  (\bibinfo  {publisher} {Oxford University Press},\ \bibinfo {address}
  {Oxford},\ \bibinfo {year} {2013})\BibitemShut {NoStop}%
\bibitem [{\citenamefont {Doi}(2015)}]{doi2015onsager}%
  \BibitemOpen
  \bibfield  {author} {\bibinfo {author} {\bibfnamefont {Masao}\ \bibnamefont
  {Doi}},\ }\bibfield  {title} {\enquote {\bibinfo {title} {Onsager principle
  as a tool for approximation},}\ }\href@noop {} {\bibfield  {journal}
  {\bibinfo  {journal} {Chin. Phys. B}\ }\textbf {\bibinfo {volume} {24}},\
  \bibinfo {pages} {1674--1056} (\bibinfo {year} {2015})}\BibitemShut {NoStop}%
\bibitem [{\citenamefont {Meng}\ \emph {et~al.}(2016)\citenamefont {Meng},
  \citenamefont {Luo}, \citenamefont {Doi},\ and\ \citenamefont
  {Ouyang}}]{MengFanlong2016a}%
  \BibitemOpen
  \bibfield  {author} {\bibinfo {author} {\bibfnamefont {Fanlong}\ \bibnamefont
  {Meng}}, \bibinfo {author} {\bibfnamefont {Ling}\ \bibnamefont {Luo}},
  \bibinfo {author} {\bibfnamefont {Masao}\ \bibnamefont {Doi}}, \ and\
  \bibinfo {author} {\bibfnamefont {Zhongcan}\ \bibnamefont {Ouyang}},\
  }\bibfield  {title} {\enquote {\bibinfo {title} {Solute based lagrangian
  scheme in modeling the drying process of soft matter solutions},}\ }\href
  {\doibase 10.1140/epje/i2016-16022-9} {\bibfield  {journal} {\bibinfo
  {journal} {Eur. Phys. J. E}\ }\textbf {\bibinfo {volume} {39}},\ \bibinfo
  {pages} {22} (\bibinfo {year} {2016})}\BibitemShut {NoStop}%
\bibitem [{\citenamefont {Di}\ \emph {et~al.}(2016)\citenamefont {Di},
  \citenamefont {Xu},\ and\ \citenamefont {Doi}}]{DiYana2016}%
  \BibitemOpen
  \bibfield  {author} {\bibinfo {author} {\bibfnamefont {Yana}\ \bibnamefont
  {Di}}, \bibinfo {author} {\bibfnamefont {Xianmin}\ \bibnamefont {Xu}}, \ and\
  \bibinfo {author} {\bibfnamefont {Masao}\ \bibnamefont {Doi}},\ }\bibfield
  {title} {\enquote {\bibinfo {title} {Theoretical analysis for meniscus rise
  of a liquid contained between a flexible film and a solid wall},}\ }\href
  {\doibase 10.1209/0295-5075/113/36001} {\bibfield  {journal} {\bibinfo
  {journal} {Europhys. Lett.}\ }\textbf {\bibinfo {volume} {113}},\ \bibinfo
  {pages} {36001} (\bibinfo {year} {2016})}\BibitemShut {NoStop}%
\bibitem [{\citenamefont {Xu}\ \emph {et~al.}(2016)\citenamefont {Xu},
  \citenamefont {Di},\ and\ \citenamefont {Doi}}]{XuXianmin2016}%
  \BibitemOpen
  \bibfield  {author} {\bibinfo {author} {\bibfnamefont {Xianmin}\ \bibnamefont
  {Xu}}, \bibinfo {author} {\bibfnamefont {Yana}\ \bibnamefont {Di}}, \ and\
  \bibinfo {author} {\bibfnamefont {Masao}\ \bibnamefont {Doi}},\ }\bibfield
  {title} {\enquote {\bibinfo {title} {Variational method for contact line
  problems in sliding liquids},}\ }\href {\doibase 10.1063/1.4959227}
  {\bibfield  {journal} {\bibinfo  {journal} {Phys. Fluids}\ }\textbf {\bibinfo
  {volume} {28}},\ \bibinfo {pages} {087101} (\bibinfo {year}
  {2016})}\BibitemShut {NoStop}%
\bibitem [{\citenamefont {Man}\ and\ \citenamefont
  {Doi}(2016)}]{ManXingkun2016}%
  \BibitemOpen
  \bibfield  {author} {\bibinfo {author} {\bibfnamefont {Xingkun}\ \bibnamefont
  {Man}}\ and\ \bibinfo {author} {\bibfnamefont {Masao}\ \bibnamefont {Doi}},\
  }\bibfield  {title} {\enquote {\bibinfo {title} {Ring to mountain transition
  in deposition pattern of drying droplets},}\ }\href {\doibase
  10.1103/PhysRevLett.116.066101} {\bibfield  {journal} {\bibinfo  {journal}
  {Phys. Rev. Lett.}\ }\textbf {\bibinfo {volume} {116}},\ \bibinfo {pages}
  {066101} (\bibinfo {year} {2016})}\BibitemShut {NoStop}%
\bibitem [{\citenamefont {Zhou}\ \emph {et~al.}(2017)\citenamefont {Zhou},
  \citenamefont {Jiang},\ and\ \citenamefont {Doi}}]{ZhouJiajia2017}%
  \BibitemOpen
  \bibfield  {author} {\bibinfo {author} {\bibfnamefont {Jiajia}\ \bibnamefont
  {Zhou}}, \bibinfo {author} {\bibfnamefont {Ying}\ \bibnamefont {Jiang}}, \
  and\ \bibinfo {author} {\bibfnamefont {Masao}\ \bibnamefont {Doi}},\
  }\bibfield  {title} {\enquote {\bibinfo {title} {Cross interaction drives
  stratification in drying film of binary colloidal mixtures},}\ }\href
  {\doibase 10.1103/PhysRevLett.118.108002} {\bibfield  {journal} {\bibinfo
  {journal} {Phys. Rev. Lett.}\ }\textbf {\bibinfo {volume} {118}},\ \bibinfo
  {pages} {108002} (\bibinfo {year} {2017})}\BibitemShut {NoStop}%
\bibitem [{\citenamefont {Man}\ and\ \citenamefont
  {Doi}(2017)}]{ManXingkun2017}%
  \BibitemOpen
  \bibfield  {author} {\bibinfo {author} {\bibfnamefont {Xingkun}\ \bibnamefont
  {Man}}\ and\ \bibinfo {author} {\bibfnamefont {Masao}\ \bibnamefont {Doi}},\
  }\bibfield  {title} {\enquote {\bibinfo {title} {Vapor-induced motion of
  liquid droplets on an inert substrate},}\ }\href {\doibase
  10.1103/physrevlett.119.044502} {\bibfield  {journal} {\bibinfo  {journal}
  {Phys. Rev. Lett.}\ }\textbf {\bibinfo {volume} {119}},\ \bibinfo {pages}
  {044502} (\bibinfo {year} {2017})}\BibitemShut {NoStop}%
\bibitem [{\citenamefont {Di}\ \emph {et~al.}(2018)\citenamefont {Di},
  \citenamefont {Xu}, \citenamefont {Zhou},\ and\ \citenamefont
  {Doi}}]{DiYana2018}%
  \BibitemOpen
  \bibfield  {author} {\bibinfo {author} {\bibfnamefont {Yana}\ \bibnamefont
  {Di}}, \bibinfo {author} {\bibfnamefont {Xianmin}\ \bibnamefont {Xu}},
  \bibinfo {author} {\bibfnamefont {Jiajia}\ \bibnamefont {Zhou}}, \ and\
  \bibinfo {author} {\bibfnamefont {Masao}\ \bibnamefont {Doi}},\ }\bibfield
  {title} {\enquote {\bibinfo {title} {Analysis of thin film dynamics in
  coating problems using {Onsager} principle},}\ }\href {\doibase
  10.1088/1674-1056/27/2/024501} {\bibfield  {journal} {\bibinfo  {journal}
  {Chin. Phys. B}\ }\textbf {\bibinfo {volume} {27}},\ \bibinfo {pages}
  {024501} (\bibinfo {year} {2018})}\BibitemShut {NoStop}%
\bibitem [{\citenamefont {Bico}\ and\ \citenamefont
  {Qu{\'e}r{\'e}}(2002)}]{bico2002rise}%
  \BibitemOpen
  \bibfield  {author} {\bibinfo {author} {\bibfnamefont {Jose}\ \bibnamefont
  {Bico}}\ and\ \bibinfo {author} {\bibfnamefont {David}\ \bibnamefont
  {Qu{\'e}r{\'e}}},\ }\bibfield  {title} {\enquote {\bibinfo {title} {Rise of
  liquids and bubbles in angular capillary tubes},}\ }\href@noop {} {\bibfield
  {journal} {\bibinfo  {journal} {J. Colloid Interface Sci.}\ }\textbf
  {\bibinfo {volume} {247}},\ \bibinfo {pages} {162--166} (\bibinfo {year}
  {2002})}\BibitemShut {NoStop}%
\bibitem [{\citenamefont
  {Princen}(1969{\natexlab{a}})}]{princen1969capillary_1}%
  \BibitemOpen
  \bibfield  {author} {\bibinfo {author} {\bibfnamefont {H.~M.}\ \bibnamefont
  {Princen}},\ }\bibfield  {title} {\enquote {\bibinfo {title} {Capillary
  phenomena in assemblies of parallel cylinders: I. capillary rise between two
  cylinders},}\ }\href {\doibase 10.1016/0021-9797(69)90379-8} {\bibfield
  {journal} {\bibinfo  {journal} {J. Colloid Interface Sci.}\ }\textbf
  {\bibinfo {volume} {30}},\ \bibinfo {pages} {69} (\bibinfo {year}
  {1969}{\natexlab{a}})}\BibitemShut {NoStop}%
\bibitem [{\citenamefont
  {Princen}(1969{\natexlab{b}})}]{princen1969capillary_2}%
  \BibitemOpen
  \bibfield  {author} {\bibinfo {author} {\bibfnamefont {H.~M.}\ \bibnamefont
  {Princen}},\ }\bibfield  {title} {\enquote {\bibinfo {title} {Capillary
  phenomena in assemblies of parallel cylinders: {II}. capillary rise in
  systems with more than two cylinders},}\ }\href {\doibase
  10.1016/0021-9797(69)90403-2} {\bibfield  {journal} {\bibinfo  {journal} {J.
  Colloid Interface Sci.}\ }\textbf {\bibinfo {volume} {30}},\ \bibinfo {pages}
  {359} (\bibinfo {year} {1969}{\natexlab{b}})}\BibitemShut {NoStop}%
\end{thebibliography}%

\end{document}